\documentstyle[preprint,revtex,eqsecnum]{aps}
\begin{document}
\tolerance 10000
\draft
\hskip 12cm
LPQTH-93/01
\vskip 1.0truecm

\begin{center}
   {\Large \bf Binding of holes and pair spectral function}
\\ {\Large \bf in the t-J model}
\end{center}
\vspace*{0.7cm}
\author{Didier POILBLANC\cite{byline}}
\vspace*{0.4cm}
\begin{instit}
\begin{center}
Laboratoire de Physique Quantique,\\
Universit\'{e} Paul Sabatier,\\
31062 Toulouse, France\cite{dbyline}
\end{center}
\end{instit}
\receipt{\hskip 3truecm}

\begin{abstract}
Clusters of the two-dimensionnal t--J model with 2 holes and up to
26 sites are diagonalized using a Lanczos algorithm. The
behaviour of the binding energy with system size suggests the existence
of a finite critical value of J above which binding occurs in the bulk.
Only the {\it d-wave} pair field operator acting on the Heisenberg GS has
a finite overlap with the 2 hole ground state for all the clusters considered.
The related spectral function associated with the propagation of a
d-wave (singlet) pair
of holes in the antiferromagnetic background is calculated.
The quasiparticle peak at the bottom of the spectrum as well as some
structure appearing above the peak survive with
increasing cluster size. Although no simple scaling law was found for
the quasiparticle weight the data strongly suggest that this weight is
finite in the bulk limit and is roughly proportional to the antiferromagnetic
coupling J (for $J<1$).
\end{abstract}
\pacs{PACS numbers: 75.10.Jm, 74.65.+n, 75.40.Mg, 74.20.-z}

\section{Introduction}
The possibility that high-temperature superconductivity in the quasi-two
dimensional cuprates would be a new phenomenon based on purely electronic
interactions \cite{review} has motivated a huge theoretical effort to
better understand the physics of two dimensional strongly correlated
electrons.
The Hubbard model and its large coupling version the t--J model are one of the
simplest theoretical models used to describe the low-energy excitations
of the copper oxide planes.
Although it is difficult to develop a satisfactory
perturbative analysis in the case of strongly repulsive interactions,
numerical methods can
often provide useful informations on both static and dynamical properties
of these systems. Exact diagonalization (ED) studies do not suffer from
random noise problems like stochastic Monte Carlo methods in particular
in the vicinity of the antiferromagnetic Mott insulator phase.
ED are therefore
well adapted to the calculations of dynamical correlation functions
at small doping density.
However, so far ED have been restricted to fairly small clusters (typically
$4\times 4$). Recently, an attempt was made to handle larger sizes (up to
26 sites) in order to perform a finite size analysis of the data
\cite{didier1,didier2,didier3} in the
close vicinity of the insulator magnetic phase.
Although this analysis was restricted to the special limit of
a single hole in the antiferromagnet (AF) it gives some hints about the
hole propagation at small but finite doping fraction \cite{didier1,didier2}
in the region where the antiferromagnetic correlation length
is still larger than the
spatial extention of the hole wavefunction. This work strongly
suggested that the quasiparticle (QP) peak
seen at the bottom of the spectral function survive in the
thermodynamic limit \cite{didier2}.
However this analysis neglects the role of a possible hole-hole effective
attraction that might occur for more than a single hole in the
AF. The purpose of the present work is then to study the case of 2 holes ie the
simplest case that nevertheless include the effect of the effective
attraction between the holes.
The issue of binding
is of great importance in the search for superconductivity
in models of strongly correlated fermions.
By a similar finite size analysis I shall attempt to give an
estimate of the binding energy and of the pair spectral function in the
thermodynamic limit.
Partial results for the binding energy and the QP weight
have been reported elsewhere \cite{binding}.
The emphasis is put here on the
energy dependence of the complete spectral function $A_{2h}(\omega)$ and on its
behavior with system size.

In standard notations the t--J Hamiltonian reads
\begin{eqnarray}
{\cal H} = &J& \sum_{{\bf i},{\vec \epsilon}}
\{(S^z_{\bf i} S^z_{\bf i+\vec \epsilon} - \frac{1}{4}
n_{\bf i} n_{\bf i+\vec \epsilon} )
+ \frac1{2}(S^+_{\bf i} S^-_{\bf
i+\vec \epsilon} + S^-_{\bf i} S^+_{\bf i+\vec \epsilon} )\}\nonumber \\
- &t& \sum_{{\bf i},{\vec \epsilon},{\sigma}}
({\tilde c}^{\dagger}_{{\bf i},{\sigma}}
{\tilde c}_{{{\bf i}+{\vec\epsilon}},{\sigma}}
+ h.c. ) .
\label {tj}
\end{eqnarray}
\noindent
The first term is the usual antiferromagnetic coupling between spins
which splits into
a diagonal part and a spin flip part. The second term describes the
hopping of the holes (ie where spins are missing).
The sum over ${\bf i},{\vec\epsilon}$ is restricted
to nearest neighbor bonds along $\vec x$ and $\vec y$ on a 2D square lattice,
${\tilde c}^{\dagger}_{{\bf i},{\sigma}}
=(1-n_{{\bf i},{-\sigma}})c_{{\bf i},{\sigma}}$ is the hole creation operator
and $n_{\bf i}=\sum_\sigma c^\dagger_{{\bf i},\sigma}c_{{\bf i},\sigma}$.
So far t will set the energy scale.

The Hamiltonian (\ref{tj}) with 0, 1 and 2 holes
is diagonalized on cluster of increasing sizes
by a standart Lanczos procedure.
It is important in this analysis that all the cluster have the same shape
namely a square. One should expect in this case a smoother extrapolation to
the thermodynamic limit. If the (integer) coordinates of the cluster periods
are $(n,m)$ and $(-m,n)$ the cluster contains $N=n^2+m^2$ sites.
The following analysis
is based on clusters of sizes $4\times 4$, $\sqrt{18}\times\sqrt{18}$,
$\sqrt{20}\times\sqrt{20}$ and $\sqrt{26}\times\sqrt{26}$.
In order to be able to
handle the largest size (26 sites) use of complete translation and
rotation symmetries (for 0 and 2 holes)
as well as time reversal (for an even number of fermions)
is necessary and the hamiltonian is diagonalized in each symmetry sector.
The size of the Hilbert space increases exponantially fast with the
cluster size. In the relevant symmetry sector (for discussion about GS symmetry
see Sec. 2) the Hilbert space for 2 holes contains after symmetry reduction
$3\,332$, $13\,858$, $58\,274$ and $4\,229\,236$ states
for $N=16$, $18$, $20$ and $26$ sites respectively.

\section{Ground state energy and binding}

Symmetry properties of the Heisenberg GS $\mid\Psi_0^N>$ and of the
two-hole (singlet) GS $\mid\Psi_0^{N-2}>$
are summarized for various cluster sizes in Tables 1A and 1B respectively.
The symmetries of the
half filled GS are exactly known \cite{marshall} but they depend
on the convention to represent the spins.
In the following I use the fermion representation
including at half filling.
$\mid\Psi_0^N>$ and $\mid\Psi_0^{N-2}>$ are both
invariant under lattice translations (total
momentum $\bf K=0$) and even (odd) under spin reversal for
4p (4p+2) fermions. This last feature is a simple consequence of the
singlet nature of these states. Although, at half filling, in the spin
language the GS is always invariant under a $90^o$ rotation
around any lattice site (s-wave) \cite{marshall},
fermion reordering can lead to an extra
minus sign, eg for $N=4$ \cite{adriana} or $N=20$ the half filled GS
becomes d-wave in the fermion convention (see Table 1A).
It is worth noticing that in such a case
the 2 hole GS is s-wave (see Table 1B).
However in most cases the half filled GS is s-wave
and the 2 hole GS is d-wave.

The physical idea of an effective attraction between the holes can
be simply understood in the Ising limit of Hamiltonian
(\ref{tj}) ie when the spin flip term is neglected. In this case the
half filled GS is the simple classical N\'eel state and obviously
two holes added in this background can minimize the magnetic energy cost
by sitting on
nearest neighbor sites. Although this crude explaination neglects the
delocalisation energy of the holes (t term) this picture was indeed shown
numerically to be correct provided that $J> 0.18$
\cite{tjz}. The situation is
not so clear in the fully quantum case (\ref{tj}) where
spin fluctuations are included.
Numerical calculations of the binding energy on
the $4\times 4$ \cite{manou,prelov1,hase,jose1},
$\sqrt{18}\times\sqrt{18}$ \cite{jose2} and
$\sqrt{20}\times\sqrt{20}$ \cite{20sites} clusters suggested that binding
still occur but above a slightly larger critical value of $J$ as expected.
{}From a recent calculation of the hole density correlation functions
in a restricted Hilbert space on clusters with up to 26 sites
Prelov\u sek and Zotos quote a critical value of $J$ of order 0.2.
However this technique gives a poor accuracy for the GS energy and can not
be applied to calculate directly the binding energy (see later).
On the other hand, Boninsegni and Manousakis
\cite{manou2} using Green Function Monte Carlo
simulations carried out on large clusters at $J\ge 0.4$ and extrapolated to
the bulk limit and smaller $J/t$ ratios report a critical value of $J$ of
order 0.28.

The GS energies of a single hole $E_0^{N-1}-E_0^N$ as well
as the corresponding
GS quantum numbers have been reported elsewhere \cite{didier1}.
The two hole GS energies $E_0^{N-2}-E_0^N$ are listed in Table 2
for various cluster sizes and J values.
I note that the size dependence is weak.
The data of the 26 site cluster confirm the $J^\nu$
dependence, $\nu\sim 0.9$, reported earlier for the $4\times 4$
cluster \cite{hase}.

The binding energy is defined quantitatively by
\begin{equation}
\Delta_{2h}=E_0^{N-2}+E_0^{N}-2 E_0^{N-1}.
\label{binding}
\end{equation}
Binding occurs when $\Delta_{2h}< 0$ in the bulk. In this case
$\mid\Delta_{2h}\mid$ gives the magnitude of the attractive
potential between the two holes. The simple (crude) arguments stated
above would give an attraction of the order of a fraction of $J$.
On the contrary, if binding does not take place one would expect
$\Delta_{2h}\rightarrow 0$ when $N\rightarrow\infty$.

$\Delta_{2h}$ vs J is shown in Fig. 1a and the corresponding data are
listed in Table 3. For $N=26$ Fig. 1a shows a significant increase of
the critical value of $J$ above which binding occurs,
$J_c\sim 0.125$ for $N=26$.
In Fig. 2a the same data are plotted vs $1/N$.
Although $\Delta_{2h}$ is a small number corresponding to
the difference between two quantities of the same magnitude
its behavior with system size is rather smooth (although not
monotonic). Note that the jump of the single hole GS
momentum \cite{didier1,remark1} between the different clusters may be
responsible for a small systematic error. Fig. 2a shows unambiguously
that binding occurs for, let say, $J> 0.5$. However
the strong size dependence at smaller $J$ does not unable to
give any accurate estimation of the critical value of $J$ for the bulk.
I simply note that even for the largest size considered here
$\Delta_{2h}\sim -J$ while one would expect a somewhat smaller
(absolute) value in the bulk. This means that these numbers
might still be far from the expected ones in the thermodynamic limit
and it is difficult then to perform any extrapolations.
On the basis of these data I can nevertheless speculate that the
critical value of J lies in the range 0.3 - 0.5. This is of central
importance since binding of holes is a necessary condition
(not sufficient) for the appearance of superconductivity.

In this work I have not studied the transition (with increasing J) towards
phase separation \cite{emery} ie when the holes tend to
cluster in a separate region of space and separate from the spins
rather than bind by pairs (if long range Coulomb repulsion is neglected).
In fact, according to previous work \cite{prelov1,xenophon} the hole
delocalization energy prevents the transition towards phase
separation to occur at the same critical value of J at which binding
first appears but rather leads to phase separation at a
larger value of J. Therefore, in this scenario, there exists a finite
range of J where holes do form pairs but do not phase separate from the
spins. Further work is clearly needed and diagonalization of 4 holes
on 26 sites would certainly help to clarify this issue. This is left for
future work.

\section{Quasiparticle weight and spectral function}

Let me now consider dynamical correlations. To study the propagation of a
pair of holes in the underlying AF background it is useful to define
the pair spectral function,
\begin{equation}
A_{2h}(\omega )=
\sum_m\mid\big<\Psi_{m}^{N-2}\mid \Delta^\dagger
\mid\Psi_0^{N}\big>\mid^2  \delta (\omega +E_0^{N}-E_m^{N-2}).
\label{spectral}
\end{equation}
As defined above $\mid\Psi_0^{N}\big>$ is the antiferromagnetic GS
at half filling (N spins on N sites).
The sum is performed over a complete set of eigenstates
$\{\Psi_{m}^{N-2}\}$ of the two hole sector (N-2 spins), with
corresponding energies $E_m^{N-2}$.
In the numerical results below, for plotting purposes,
the $\delta$--functions are replaced by sharp Lorentzians of small width
$\epsilon \simeq 0.02$.
$\Delta^\dagger$ is the usual pair creation operator,
\begin{equation}
\Delta^\dagger=\frac{1}{\sqrt{N}}\sum_{\bf i,\vec\epsilon} {\cal
F}({\vec\epsilon})
\, {\tilde c}^{\dagger}_{{\bf i},{\uparrow}}
{\tilde c}^{\dagger}_{{{\bf i}+{\vec\epsilon}},{\downarrow}} .
\label{pair}
\end{equation}
where the form factor ${\cal F}(\vec\epsilon)$ ($=\pm 1$) can have s- or d-wave
symmetry. As seen previously, it is remarkable that the half filled and
two hole GS have systematically opposite quantum numbers (ie $\pm 1$) under
a $90^o$ rotation (around a lattice site). Therefore
it is clear that only
the {\it d-wave} pair operator (ie ${\cal F}(\pm \vec x)=1$
and ${\cal F}(\pm \vec y)=-1$) has a non-zero matrix element
between the two GS. For this reason, I shall restrict myself to
this choice so far. Note that the pair operator (\ref{pair}) is odd under spin
reversal and is invariant under lattice translations so that, in addition
to its d-wave nature, it has precisely
the right translation and spin symmetries (it is a singlet)
to connect the two GS. In the right hand side of
Eq. (\ref{spectral}) the sum is then restricted to the states of the same
symmetry class as the 2 hole GS.

The calculation of the spectral function (\ref{spectral}) is done
in three steps. First the Heisenberg GS is calculated.
In practice the number of Lanczos iterations required to obtain a good
accuracy of the GS increases slowly with the system size (crudely it
is proportional to the system size). Typically, for the Heisenberg GS I
have reached
a high accuracy after 60, 150, 400 and 600 iterations for $N=16$,
$18$, $20$ and $26$ respectively. Note that the GS energy
converges in much less iterations than the corresponding wavefunction,
in any case less than 50 iterations.
In the second step the pair operator is
applied on the Heisenberg GS. Lastly the vector generated in the second step
is used as the initial vector of a second Lanczos run in the 2
hole subspace. The spectral function can be generated by a continued
fraction expension that can be truncated after $N_{it}$ iterations
\cite{contfrac}.
I have carefully studied the convergence of the spectral function as a
function of $N_{it}$. Fig. 3 shows the behavior of $A_{2h}$ for
increasing $N_{it}$ in the case of the 26 site cluster. At small $N_{it}$
the spectral weight is distributed on a small number of $\delta$-functions.
It is important to note that the low energy part of the spectrum is the first
to converge when $N_{it}$ is increased. Fig. 3 shows that a complete
convergence on the whole physically relevant energy range
can be obtained after a rather small number of iterations much smaller than
the size of the Hilbert space. Empirically I have noticed that the value of
$N_{it}$ required to obtain a good convergence does not grow much faster
than, let say, the system size N. For example, $N_{it}=200$, 400,
500 and 450 iterations are needed for 16, 18, 20 and 26 sites.

The spectral function is plotted in Fig. 4 for increasing cluster size
and for $J=0.1$, $0.3$ and $1$. I observe that most of the
features of the $4\times 4$
cluster \cite{pairsuscep} survive when the cluster size is increased;
(i) a QP peak lies at the bottom of the band and (ii) a significant
amount of spectral weight is spread at higher energy over a range of a few t
(for details see discussion below).
This is very similar to the single hole spectral function
\cite{didier1,didier2}. I note that the few peaks seen in the case
of the smallest size (16 sites) are smeared out when N is increased
and eventually merge into
an almost perfect continuous background for N=26 except for $J=1$.
It is also remarkable that the various local
maxima in the background (at $\omega\sim 0.8$ in Fig. 4d or
$\omega\sim 0.5$ , $2$ and $4$ in Fig. 4h)
are located at almost the same frequencies for the
various sizes N considered. However it is not clear whether the secondary peaks
(of finite width) observed immediately above the QP peak in
Figs. 4h and 4l are still finite
size effects or whether they are related to any physical low energy resonances
(see later for discussion).
It should be noted that, similarly to the
single hole case \cite{didier3}, Figs. 4 suggest that no real gap separates the
QP peak from the rest of the spectrum at higher energy.

Let me now study more quantitatively the behavior of the weight of the
QP peak. Similarily to the one-hole case, I can define the Z-factor
as the square of the overlap \cite{bob},
\begin{equation}
Z_{2h}=\frac{
\mid\big<\Psi_{0}^{N-2}\mid \Delta^\dagger
\mid\Psi_0^{N}\big>\mid^2 }{
\big<\Psi_{0}^{N}\mid \Delta \Delta^\dagger
\mid\Psi_0^{N}\big>} .
\label{z2h}
\end{equation}
The denominator is a simple normalization factor \cite{note} so that
$Z_{2h}$ represents the {\it relative} spectral weight (between
0 and 1) located in the QP peak.

One of the key issues of the problem of strongly correlated fermions
is the possibility that spin and charge decoupling
which takes place in one dimension \cite{haldane} would also occur
in two dimensions as proposed by Anderson \cite{phil} and
others \cite{marginal}. If such a scenario is valid then, as a simple
consequence of the spin-charge separation, one would expect the absence
of sharp QP. It is then crutial to test this hypothesis.
Previous calculations \cite{didier1,didier2} suggested that
the single hole Green function in the AF background does exhibit an undumped
QP pole. However, since this approach neglected any possible hole-hole
interaction it is necessary to re-examine the same problem when
more than a single hole is present. The case of two holes considered here
is the first step towards the finite hole density.

$Z_{2h}$ is plotted in Fig. 1b as a function of J. Fig. 1b shows that
$Z_{2h}$ varies almost linearly with J in the range $0.1< J < 1$.
It is also interesting to compare these data with the
single hole QP weight $Z_{1h}$ obtained recently on the same
clusters \cite{didier2},
\begin{equation}
Z_{1h}=\sum_\sigma\mid\big<\Psi_{0,\sigma}^{N-1}\mid
{\tilde c}_{\bf K,\sigma}^\dagger\mid\Psi_0^{N}\big>\mid^2 ,
\label{zfactor}
\end{equation}
$\bf K$ stands for the finite momentum of the
one hole GS \cite{remark1}. Table 4 shows the data for $Z_{2h}$ for
the various clusters as well as $(Z_{1h})^2$ on 26 sites.
We note that, to a good approximation, $Z_{2h}\simeq (Z_{1h})^2$.

Although $Z_{2h}$ is quite small in the physically interesting range
$0.1<J<0.5$ only a finite size analysis can tell whether or not it
vanishes in the bulk. The data for $Z_{2h}$ are plotted in Fig. 2b as
a function of $1/N$. It should be noted that, contrary to the
one hole case \cite{didier2,remark1} the GS momentum is identical for
all the clusters and equal to ${\bf 0}$.
No simple scaling law can be deduced from these
plots. However, the numbers at N=26 seem to indicate a convergence towards
a finite number in the bulk.

Lastly, let me consider the behavior of the spectral function vs the coupling
constant J as shown in Figs. 5a-f.
There are many similarities with the behavior of the single hole
spectral function calculated on the $4\times 4$ \cite{singlehole} and
larger clusters \cite{didier3}.
In the limit J=0 the spectrum in Fig. 5a is symmetric around $\omega =0$.
This is in fact easy to understand. The change $\omega\rightarrow -\omega$
is equivalent to $t\rightarrow -t$ ie ${\tilde c}_{{\bf i},\sigma}
\rightarrow -{\tilde c}_{{\bf i},\sigma}$ for, let say, the even sites.
Under such a transformation one can easily check that
$\Delta^\dagger\rightarrow -\Delta^\dagger$ so that eventually
$A_{2h}(\omega)=
A_{2h}(-\omega)$ for J=0. The spectral density is almost constant
over an energy range of $\sim 13\, t$. A sharp peak (actually a
$\delta$-function) can be seen at $\omega=0$.
When J is turned on (Figs. 5b,c) the spectrum is no longer symmetric
with respect to $\omega$ and no longer (rigourously) bounded
from above (strictly speaking in the bulk) since there
now exist states with excitation up to
a magnetic energy $\propto N\, J$, N being the system size.
However, in pratice, the matrix elements in (\ref{spectral})
decrease exponantially fast with increasing energy outside
an energy range of a few t ($\sim$ 10 -- 12 t).
As seen above a QP peak of weight $\propto J$ appears at the bottom of the
band. However for small J most of the spectral weight lies above.
The sharp $\omega=0$ peak of the $J\rightarrow 0$ limit is rapidly broaden by
J.
Reminiscence of this peak can still be clearly seen up to J=0.2.
If J is increased further (figs. 5d-f) most of the features of the
$J=0$ limit vanish. On the other hand, sharp peaks (nevertheless
of finite width) appear right above the QP peak. These structures are
characteristic of the $J\simeq t$ region and might be related to the
internal structure of the pair. Indeed,
the 2 holes (for sufficiently large J)
are confined on neighboring sites by some effective potential which naively
increases linearly with the hole separation (if spin fluctuations are
not considered). It is interesting to note
that the appearance of these resonances coincide with the approximate
J value at which binding sets up (see Sec. II).

\section{Conclusion}

In this paper, calculations of the binding energy as well as of the
pair field spectral function (of d-wave symmetry) have been reported
for the t--J model.
By diagonalising clusters with up to 26 sites and 2 holes it has been
definitely established that binding between holes occurs only above
a critical threshold of J in agreement with previous work. I estimate
this critical value to be in the range 0.3 -- 0.5.
The spectral function of the pair operator shows unambiguously a QP
peak at the bottom of the band which does not seem to
broaden or disappear when the system size is increased.
Its weight grows almost like J for $J<1$.
In the physical region $0.1<J<0.5$ most of the spectral weight
is spread above the peak on a wide energy range of the order of a few t.
Two qualitatively different regimes are found:
(i) for $J\le 0.25$, the spectral function exhibits a pronounced bump
around the band center which eventually merges into a sharp $\omega=0$ peak
when J=0; (ii) for $J> 0.25$, especially $J\sim 1$,
sharp (small) resonances are seen above the QP peak.
These structures might be related to the internal nature of the pair
bound by some effective potential.

The numerical calculations were done on the CRAY--2 of Centre de
Calcul Vectoriel pour la Recherche (CCVR), Palaiseau, France.
I am particularly greatful to the CCVR staff for helpful
assistance. I also wish to thank the Supercomputer Computation
Research Institute (Tallahassee) for hospitality at an early
stage of this work and acknowledge my co-workers
E. Dagotto, J. Riera, A. Moreo and M. Novotny
for many fruitful discussions. I am also indebted to J. Riera for providing me
with many data to check the computer codes.

\newpage
\hsize=17truecm
\vglue 0.3truecm
\centerline{TABLE 1A}
\vskip 1truecm

$$\vcenter{\tabskip=0truecm\offinterlineskip
\def\tablerule{\noalign{\hrule}}
\halign to17truecm{\strut#& \tabskip=0.2em plus0.2em&
  \hfil#\hfil& \vrule#&
  \hfil#\hfil&
  \hfil#\hfil&
  \hfil#\hfil& \hfil#\hfil&
  \hfil#\hfil& \hfil#\hfil&
  \hfil#\hfil\tabskip=0pt\cr
& && $4$ & $8$ & $10$ & $16$ & $18$
& $20$ & $26$ \cr
\tablerule
\tablerule
& $S_z\rightarrow -S_z$ && $+1$ & $+1$ & $-1$
& $+1$ & $-1$ & $+1$ & $-1$ \cr
& $\bf K$ && $(0,0)$ & $(0,0)$ & $(0,0)$
& $(0,0)$ & $(0,0)$ & $(0,0)$ & $(0,0)$ \cr
& ${\cal R}_{90^o}$ && d & s & s
& s & s & d & s \cr
}}
$$

\vskip 2.0truecm
\centerline{TABLE 1B}
\vskip 0.6truecm

$$\vcenter{\tabskip=0truecm\offinterlineskip
\def\tablerule{\noalign{\hrule}}
\halign to17truecm{\strut#& \tabskip=0.2em plus0.2em&
  \hfil#\hfil& \vrule#&
  \hfil#\hfil& \hfil#\hfil&
  \hfil#\hfil& \hfil#\hfil&
  \hfil#\hfil\tabskip=0pt\cr
& && $10$ & $16$ & $18$
& $20$ & $26$  \cr
\tablerule
\tablerule
& $S_z\rightarrow -S_z$ &&  $+1$ & $-1$
& $+1$ & $-1$ & $+1$ \cr
& $\bf K$ && $(0,0)$
& $(0,0)$ & $(0,0)$ & $(0,0)$ & $(0,0)$ \cr
& ${\cal R}_{90^o}$ && d
& d & d & s & d \cr
}}
$$

\newpage
\hsize=17.5truecm
\vglue 2.5truecm
\centerline{TABLE 2}
\vskip 1.2truecm

$$\vcenter{\tabskip=0truecm\offinterlineskip
\def\tablerule{\noalign{\hrule}}
\halign to17.5truecm{\strut#& \tabskip=0.2em plus0.2em&
  \hfil#\hfil& \vrule#&
  \hfil#\hfil&
  \hfil#\hfil&
  \hfil#\hfil& \hfil#\hfil&
  \hfil#\hfil& \hfil#\hfil&
  \hfil#\hfil\tabskip=0pt\cr
& && $0.1$ & $0.15$ & $0.2$ & $0.3$ & $0.5$
& $0.75$ & $1$ \cr
\tablerule
\tablerule
& $16$ && $-5.211929$ & $-4.761848$ & $-4.366113$
& $-3.651412$ & $-2.370082$ & $-0.922098$ & $0.422346$ \cr
& $18$ && $-5.372725$ & $-4.989488$ & $-4.622850$
& $-3.925908$ & $-2.635292$ & $-1.154138$ & $0.227037$ \cr
& $20$ && $-5.391175$ & $-5.002799$ & $-4.632916$
& $-3.932952$ & $-2.644911$ & $-1.174090$ & $0.195934$ \cr
& $26$ && $-5.293633$ & $-4.873587$ & $-4.489433$
& $-3.786540$ & $-2.527792$ & $-1.099776$ & $0.235903$ \cr
}}
$$

\newpage
\hsize=17.5truecm
\vglue 2.5truecm
\centerline{TABLE 3}
\vskip 1.2truecm

$$\vcenter{\tabskip=0truecm\offinterlineskip
\def\tablerule{\noalign{\hrule}}
\halign to17.5truecm{\strut#& \tabskip=0.2em plus0.2em&
  \hfil#\hfil& \vrule#&
  \hfil#\hfil&
  \hfil#\hfil&
  \hfil#\hfil& \hfil#\hfil&
  \hfil#\hfil& \hfil#\hfil&
  \hfil#\hfil\tabskip=0pt\cr
& && $0.1$ & $0.15$ & $0.2$ & $0.3$ & $0.5$
& $0.75$ & $1$ \cr
\tablerule
\tablerule
& $16$ && $-0.121909$ & $-0.133874$ & $-0.169305$
& $-0.257734$ & $-0.437502$ & $-0.661252$ & $-0.887262$ \cr
& $18$ && $0.016983^*$ & $-0.098624^*$ & $-0.194140$
& $-0.323806$ & $-0.535906$ & $-0.757922$ & $-0.988101$ \cr
& $20$ && $-0.036409$ & $-0.115937$ & $-0.182936$
& $-0.295488$ & $-0.4798095$ & $-0.681616$ & $-0.871744$\cr
& $26$ && $0.059531$ & $$ & $-0.051799$
& $-0.154960$ & $-0.340200$ & $$ & $-0.738611$\cr
}}
$$

\newpage
\hsize=17.5truecm
\vglue 2.5truecm
\centerline{TABLE 4}
\vskip 1.2truecm

$$\vcenter{\tabskip=0truecm\offinterlineskip
\def\tablerule{\noalign{\hrule}}
\halign to17.5truecm{\strut#& \tabskip=0.2em plus0.2em&
  \hfil#\hfil& \vrule#&
  \hfil#\hfil&
  \hfil#\hfil&
  \hfil#\hfil& \hfil#\hfil&
  \hfil#\hfil& \hfil#\hfil&
  \hfil#\hfil\tabskip=0pt\cr
& && $0.1$ & $0.15$ & $0.2$ & $0.3$ & $0.5$
& $0.75$ & $1$ \cr
\tablerule
\tablerule
& $16$ && $0.017334$ & $0.043994$ & $0.069076$
& $0.112460$ & $0.193285$ & $0.286379$ & $0.376510$ \cr
& $18$ && $0.028308$ & $0.039858$ & $0.052372$
& $0.080068$ & $0.142729$ & $0.225824$ & $0.306358$ \cr
& $20$ && $0.011259$ & $0.027350$ & $0.038865$
& $0.066018$ & $0.130332$ & $0.214760$ & $0.294387$\cr
& $26$ && $0.014675$ & $0.031535$ & $0.050113$
& $0.089192$ & $0.164575$ & $0.248636$ & $0.322983$\cr
& $26^*$ && $$ & $$ & $$
& $0.081419$ & $0.155884$ & $$ & $0.318740$\cr
}}
$$

\newpage
\hsize=17truecm
\centerline {TABLE CAPTIONS}
\vskip 2truecm

\noindent
{\bf Table 1}

\noindent
Symmetry properties of the half filled (A) and two hole (B) GS.
The fermion convention is used.
\bigskip

\noindent
{\bf Table 2}

\noindent
Two hole ground state energies of the t--J model $E_0^{N-2}-E_0^N$
on clusters $4\times 4$,
$\sqrt{18}\times\sqrt{18}$, $\sqrt{20}\times\sqrt{20}$ and
$\sqrt{26}\times\sqrt{26}$ for several values of J (0.1 to 1).
\bigskip

\noindent
{\bf Table 3}

\noindent
Binding energies in the t--J model on clusters $4\times 4$,
$\sqrt{18}\times\sqrt{18}$, $\sqrt{20}\times\sqrt{20}$ and
$\sqrt{26}\times\sqrt{26}$ for several values of J (0.1 to 1).
Note that for 18 sites a level crossing appears in the
one hole sector (indicated by $^*$).
\bigskip

\noindent
{\bf Table 4}

\noindent
Spectral weight $Z_{2h}$ of the QP peak in the pair spectral function
of the t--J model on clusters $4\times 4$,
$\sqrt{18}\times\sqrt{18}$, $\sqrt{20}\times\sqrt{20}$ and
$\sqrt{26}\times\sqrt{26}$ for several values of J (0.1 to 1).
For comparison, $Z_{1h}^2$ corresponding to the one hole GS on 26 sites
is also listed on the last line (marked by $^*$).
\bigskip

\newpage
\centerline {FIGURE CAPTIONS}
\vskip 2truecm

\noindent
{\bf Figure 1}

\noindent
Binding energy $\Delta_{2h}$ (a) and spectral weight $Z_{2h}$ (b) vs J.
The symbols corresponding to the different clusters are shown on the plot.
\bigskip

\noindent
{\bf Figure 2}

\noindent
Binding energy $\Delta_{2h}$ (a) and spectral weight $Z_{2h}$ (b) plotted
(arbitrarily) vs inverse system size $1/N$.
The value of J corresponding to each set of points is indicated on the plot
on the left side of the corresponding data.
\bigskip

\noindent
{\bf Figure 3}

\noindent
Spectral function of the d-wave pair operator vs frequency (in unit of t) for
$N=26$ and $J=0.3$. The Lanczos iteration number $N_{it}$ used to calculate
the continued fraction of the spectral function is shown on the plots.
\bigskip

\noindent
{\bf Figure 4}

\noindent
Spectral function of the d-wave pair operator vs frequency for
increasing cluster sizes. $J=0.1$ (a -- d), $0.3$ (e -- h) and $1$
(i -- l). The frequency
$\omega$ is measured in unit of t.
\bigskip

\noindent
{\bf Figure 5}

\noindent
Spectral function of the d-wave pair operator vs frequency on the
$\sqrt{26}\times\sqrt{26}$ cluster for
various J values. $\omega$ is measured in unit of t.
The continued fraction expension was truncated after $N_{it}=450$
iterations (except for $J=0$ where $N_{it}=550$).
\bigskip


\begin{thebibliography}{10}
\bibitem[*]{byline} E-mail: didier@occitana.ups-tlse.fr

\bibitem[**]{dbyline} Laboratoire Associ\'e au Centre National de la
Recherche Scientifique (URA 505)

\bibitem{review} P. W. Anderson and J. R. Schrieffer,
Physics Today {\bf 44}, no 6, pp 54-61 (1991).

\bibitem{didier1} D. Poilblanc, H. J. Schulz, and T. Ziman,
Phys. Rev. B {\bf 46}, 6435 (1992).

\bibitem{didier2} D. Poilblanc, H. J. Schulz, and T. Ziman,
Phys. Rev. B {\bf 47}, 3273 (1993).

\bibitem{didier3} D. Poilblanc, T. Ziman, H. J. Schulz and
E. Dagotto, Phys. Rev. B, in press (1993).

\bibitem{binding} D. Poilblanc, J. Riera, and E. Dagotto,
submitted to Phys. Rev. Letters.

\bibitem{marshall} W. Marshall, Proc. R. Soc. London A{\bf 232}, 48 (1955).

\bibitem{adriana} For clusters of size $N=M\times M$,
the half filled GS of the Hubbard or Heisenberg model
is, in the fermion representation,
s-wave when M/2 is even and d-wave when  M/2 is odd. See A. Moreo and
E. Dagotto, Phys. Rev. B{\bf 41}, 9488 (1990).

\bibitem{tjz} J. Riera and E. Dagotto, SCRI--FSU preprint.

\bibitem{manou} E. Kaxiras and E. Manousakis, Phys. Rev. B{\bf 38},
566 (1988).

\bibitem{prelov1} J. Bon\c ca, P. Prelov\u sek, and I. Sega, Phys.
Rev. B{\bf 39}, 7074 (1989).

\bibitem{hase} Y. Hasegawa and D. Poilblanc,
Phys. Rev. B{\bf 40}, 9035 (1989).

\bibitem{jose1} J. Riera, Phys. Rev. B{\bf 40}, 833 (1989).

\bibitem{jose2} J. Riera, Phys. Rev. B{\bf 43}, 3681 (1991).

\bibitem{20sites} T. Itoh, M. Arai, and T. Fujiwara,
Phys. Rev. B{\bf 42}, 4834 (1990);
H. Fehske, V. Waas, H. R\"oder and H. B\"uttner,
Phys. Rev. B{\bf 44}, 8473 (1991).

\bibitem{xenophon} P. Prelov\u sek and X. Zotos,
Phys. Rev. B, in press.

\bibitem{manou2} M. Boninsegni and E. Manousakis, FSU-SCRI preprint (1992).

\bibitem{remark1} For most clusters (except $4\times 4$) the one
hole momentum $(\pi/2,\pi/2)$ does not belong to the reciprocal space
of the cluster so that the actual GS momentum is the allowed wavevector
in the closest vicinity of $(\pi/2,\pi/2)$.

\bibitem{contfrac} R. Haydock, V. Heine, and M. J. Kelly, J. Phys. C{\bf 8}
2591 (1975); E. R. Gagliano and C. A. Balseiro, Phys. Rev. Lett. {\bf 59}
2999 (1987).

\bibitem{emery}  V. Emery, S. Kivelson and H. Q. Lin,
Phys. Rev. Lett. {64}, 475 (1990).

\bibitem{pairsuscep} E. Dagotto, J. Riera and A. P. Young,
Phys. Rev. B{\bf 42}, 2347 (1990).

\bibitem{bob} E. Dagotto and J. R. Schrieffer, Phys. Rev. B{\bf 43}, 8705
(1991).

\bibitem{note} This normalization factor is equal to the integral of
the spectral function over $\omega$. It is almost independent of
system size. I found $2.403\, 56$, $2.387\, 97$,
$2.381\, 62$ and $2.368\, 90$ for $N=16$, $18$, $20$
and $26$ respectively.

\bibitem{haldane} F. D. M. Haldane, J. Phys. C. {\bf 14}, 2585 (1981).

\bibitem{phil} P. W. Anderson, Phys. Rev. Lett. {\bf 64}, 1839 (1990).

\bibitem{marginal} C. M. Varma et al., Phys. Rev. Lett. {\bf 63}, 1996 (1989).

\bibitem{singlehole} E. Dagotto, R. Joynt, A. Moreo, S. Bacci and
E. Gagliano, Phys. Rev. B{\bf 41}, 9049 (1990).

\end{thebibliography}
\end{document}